\RequirePackage{fix-cm}
\documentclass[smallextended]{svjour3}       
\smartqed  
\usepackage{amssymb}
\usepackage{amsmath}
\usepackage{graphicx}
\begin{document}

\title{Travelling waves of  density   for a fourth-gradient model of fluids   
}
\subtitle{}

\titlerunning{Waves of  density   for a fourth-gradient model}        

\author{Henri Gouin  $^{(a)}$ \\  Giuseppe  Saccomandi $^{(b)}$
}

\institute{ \at $^{(a)}$ Corresponding author\\
             Aix-Marseille Universit\'e, CNRS, Centrale Marseille, M2P2 UMR 7340,\ \ 13451  Marseille, France \\
              \email{henri.gouin@univ-amu.fr; henri.gouin@yahoo.fr}
\\
            $^{(b)}$  Dipartimento di Ingegneria,   Universit\`{a} degli Studi di Perugia,
          06125  Perugia, Italy \\
              \email{giuseppe.saccomandi@unipg.it} }

 \date{Received: 2015-10-17 / Accepted: 20-01-2016}


\maketitle

\noindent The final publication is available at Springer via\\ http://dx.doi.org/10.1007/s00161-016-0492-3,\\

\begin{abstract}In  mean-field  theory, the non-local state  of fluid molecules can be taken into account using a statistical method.   The molecular model combined with  a density expansion in Taylor series of the fourth order  yields  an internal energy value relevant to the fourth-gradient model, and  the equation of isothermal motions takes then  density's spatial derivatives  into account    for waves travelling in both liquid and vapour phases. At  equilibrium, the equation of the density profile  across  interfaces is more precise than the \emph{Cahn and Hilliard equation}, and near the fluid's critical-point, the density profile verifies an \emph{Extended Fisher-Kolmogorov  equation}, allowing kinks,  which converges towards the  Cahn-Hillard equation  when approaching the  critical point. Nonetheless, we also get pulse waves oscillating and generating critical opalescence.
\end{abstract}

\section{Introduction}

In regions where  mass density $\rho $ is not uniform, the van der Waals
forces exert stresses on fluid molecules, producing surface tension effects \cite{Lifshitz,Evans,Widom}.
The mean-field molecular theory generates a system of  tensions \cite{rowlinson}, and its main continuous representation is known as  the second-gradient
model  \cite{Germain1}. The second-gradient model can provide  a construction of both the
free energy density of the  form $F\left(\rho , ({\rm{grad}}\rho)^2  ,T \right)$ and the internal energy density of the  form $\varepsilon \left(\rho ,({\rm{grad}}\rho)^2 ,\eta\right)$, as temperature $T$ and specific
entropy $\eta $ are conjugate variables  and the two energy densities are each the
Legendre  transformation   of the other; for fluids,  Cahn and Hilliard \cite{Cahn} demonstrated that
 $F$ and $\varepsilon $ can be assumed to be function of $\Delta\,\rho $ instead of $({\rm{grad}}\rho)^2$, where $\Delta$ and $ \, {\rm{grad}} $ are the  harmonic operator and the gradient, respectively.

The use of higher-order strain-gradient models has been studied for a long time in solid mechanics (\cite{Maugin,Rosenau,Dell'Isola1,Gouin10} and references therein). One of the reason, because higher gradients are introduced in  the framework of classical continuum mechanics, is to introduce dispersive effects in the mathematical model. This is a necessary step to study dispersive wave or localisation phenomena \cite{Peerlings,Askes}.
In fluid dynamics mathematical models that takes into account higher order derivatives of the stretching tensor have been introduced mainly to describe dipolar fluids and turbulence \cite{Dipolar,Rubin,Gurtin,Jordan}.
In the present paper we consider higher gradients of the density. This is the strategy used by Korteweg to study capillarity \cite{TN}, in the framework of the van der Waals model in the critical region \cite{Gouin3}, and to study fluid mechanics in nanotubes \cite{GS}.

It is interesting to notice that the study of models containing higher order derivatives of the density has a clear interpretation in the framework of the mean-field molecular model.
In fact, the extension obtained via the request of molecular range turns out to be effective in the construction of a new interpolating model compatible with fluctuations of density near the critical point. The internal energy takes the fourth derivative of the density into account in a three-dimensional space. The tension-like coefficients are not just formal computation: their values can be estimated by molecular forces and compared near the critical point.

Starting from the classical framework of  kinetic theory of gases  \cite{Rocard}, and using as basic constitutive quantities the potentials of the  van der Waals
forces, as done in \cite{Israel,Gouin1}, we consider an expansion in the density up to the fourth order and we obtain a new model for the volume energy.
This new model is named  a \emph{fourth-gradient} fluid.
We point out that the truncation of the Taylor's expansion to the fourth order is dictated
by the fact that if we use the principle of virtual works (or virtual powers) and we consider that at the boundary edges and points of the domain of interest only vector forces are applied,
this is the higher meaningful derivative in the sense of distributions theory \cite{Schwartz,Gouin4}.
The physical motivation is given by the interest to have a more detailed model of the interface at the nanoscale where small oscillations on the density profile are observed \cite{simulation}.

Using this approach, the equation of density profile  through  planar fluid interfaces is deduced together with the equation of isothermal travelling waves of density in liquid and vapour phases.
Near the critical point, the  equation of fluid density  for equilibrium and motion is in the form of an Extended Fisher-Kolmogorov   equation \cite{Swift}.
In the case of water,  it is possible to estimate the relative values of the constitutive coefficients
and consequently to compare the solutions of the Extended Fisher-Kolmogorov equation with the ones coming from the Cahn and Hilliard  equation. It is interesting
that this new model is able to predict very interesting solutions that share important feature with what is observed in experiments at least from a qualitative point of view \cite{simulation}.

The paper is organised as follows.   In Section \ref{sec2}, we present a three-dimensional molecular model issued from the mean-field theory with application to water in the case of London's forces.
In Section \ref{sec3}, we propose the conservative equation of isothermal motions. In Section \ref{sec4}, we consider   planar interfaces which are specific to the  vicinity of fluid's critical point; the profile of density near the fluid's critical point verifies an Extended Fisher-Kolmogorov equation generating oscillating pulses of density in the case of isothermal travelling waves. A conclusion and an appendix with some technical details end  the paper.

\section{\label{sec2} A fourth-gradient fluid energy}

\subsection{\label{sec2.1}General case}
In the  mean-field theory of hard-sphere molecules, all the fluid's
molecules  are identical   and of  mass $m_{l}$.
The central forces between molecules  derive from a potential denoted $\varphi
(r)\equiv m_{l}^{2}\psi (r)$, where $r$ is the distance between the centres  of two molecules.
\newline
In three-dimensional  Euclidian medium  $\mathcal{D}$,  the potential energy $W_{_O}$ resulting from the combined action of all the
molecules  on the molecule located at origin $O$ is assumed to be additive such that,
\begin{equation*}
W_{_O}= \sum_i m_l^2 \psi (r_i).
\end{equation*}
The summation is extended to all the fluid's molecules (except for the molecule located at origin $O$) and $r_i$ is the distance between the centres of molecule $i$  and of the molecule at $O$.
The number of molecules in the volume $dv$ is represented by $\nu(x,y,z)\, dv$, where $dv$ denotes the volume element in $\mathcal{D}$ at point of coordinates $x, y, z$, and in a continuous representation,
\begin{equation*}
W_{_O}=\iiint_{\mathcal{D}}\varphi (r)\,\nu \,dv\equiv \int_{\sigma _{l}}^{\infty
}\varphi (r)\left[ \iint_{S}\nu \,ds\right] dr,
\end{equation*}%
where $S$ is the sphere of centre $O$ and radius $r$, and $\sigma _{l}$ is
the molecular diameter. We assume that $\nu$ is an analytic
function of coordinates $x, y, z$, i.e.
\begin{equation} \label{Expansion}
\nu =\nu(0,0,0)
+\sum_{n=1}^{\infty }\frac{1}{n\,!}\left[ \,x\frac{\partial \nu
}{\partial x}(0,0,0)+y\frac{\partial \nu }{\partial y}(0,0,0)+z\frac{%
\partial \nu }{\partial z}(0,0,0)\,\right] ^{\,(n)}.   \end{equation}
Usually  expansion (\ref{Expansion}) is limited to the second order. We notice that
for any integers $p,q,r$,
\begin{equation*}
\iint_{S}x^{2p+1}y^{q}\,z^{r}ds=0,
\end{equation*}%
and
\begin{equation*}
\iint_{S}x^{2}ds=\iint_{S}y^{2}ds=\iint_{S}z^{2}ds=\frac{4\,\pi \,r^{4}}{3}.
\end{equation*}%
Then,
\begin{equation*}
W_{_O}=\int_{\sigma _{l}}^{\infty }m_{l}^{2}\psi (r)\left[ 4\,\pi \,r^{2}\nu
_{_O}+\frac{2\pi }{3}\,r^{4}\Delta \nu _{_O}\right] dr.
\end{equation*}%
Here $\nu
_{_O} \equiv \nu
 (0,0,0)$ and $\Delta \nu _{_O}\equiv \Delta \nu (0,0,0)$. Let us denote
\begin{equation}
2\,\kappa = \int_{\sigma _{l}}^{\infty }4\,\pi \,r^{2}\psi (r)\,dr, \qquad 2\,k\
b^{2} = \int_{\sigma _{l}}^{\infty }\frac{2\,\pi }{3}\,r^{4}\psi (r)\,dr,
\label{lambda} \end{equation}%
where $b$\ \ is  the fluid's  covolume  \cite{Rocard}. Then,
\begin{equation*}
W_{_O}=2\,m_{l}^{2}k\left[ \nu _{_O}+b^{2}\Delta\nu _{_O}\right]=2\,m_{l}k\left[ \rho _{_O}+b^{2}\Delta\rho _{_O}\right],
\end{equation*}
where $\rho _{_O}=m_{l}\nu _{_O}$ is the mass density at $O$.  Consider that
couples of molecules are counted twice, the potential energy density per unit volume is
\begin{equation*}
E_{_O}=\frac{1}{2}\,\nu _{_O}W_{_O} =  k\left[ \rho _{_O}^{2}+b^{2}\rho
_{_O}\Delta\rho _{_O}\right]
\end{equation*}%
and the corresponding potential energy of all the fluid is
\begin{equation}
W=\iiint_{\mathcal{D}}k\left[ \rho ^{2}+b^{2}\rho \,\Delta\rho \right] dv .\label{2grad}
\end{equation}%
To obtain the internal energy, we have to take into account  the
kinetic effects of molecular motions where   first term $k\,\rho ^{2}$  in Eq. (\ref{2grad})
corresponds to the internal pressure. Consequently, the specific
internal energy writes
\begin{equation*}
\varepsilon =\alpha (\rho ,\eta )+{k\,b^{2}}\,\rho\, \Delta\rho,
\end{equation*}
where $ \alpha (\rho ,\eta )$ is the internal energy of the homogeneous fluid of densities $\rho , \eta $.
But,
\begin{equation*}
\rho \,\Delta\rho    = \rm{div}(\rho\,{\rm{grad}}\,\rho
)-({\rm{grad}}\,\rho )^{2},
\end{equation*}%
where\ $\rm{div}$\ is the divergence operator and $\Delta \rho \equiv \rm{div} {\rm{grad}}\,\rho    $.
\begin{equation*}
W=\iiint_{\mathcal{D}}k\left[ \rho ^{2}-b^{2}({\rm{grad}}\rho )^{2}\right]
\,dv+\iint_{\Sigma }k\,b^{2}\rho \ (\boldsymbol{n}.{\rm{grad}}\,\rho)
\,d\sigma,
\end{equation*}%
where $\boldsymbol{n}$ is the external unit vector to $\Sigma$. When we assume that $\rho $ is
uniform on the boundary, the flux term on boundary $\Sigma $ is null. If we note $\lambda =- 2\, k\, b^{2}$, we get the
internal energy per unit volume in the gradient form :
\begin{equation*}
\rho \,\varepsilon =\rho \, \alpha (\rho ,\eta )+\frac{\lambda }{2}\,({\rm{grad}}\,\rho )^{2},
\end{equation*}%
which corresponds to the model of internal capillarity for the simplest case of
second-gradient theory \cite{Widom,Gouin1,Gouin4}.

Now we consider an expansion of  Eq. (\ref{Expansion}) up to the fourth order.
Odd order terms have zero integrals,  then
\begin{equation*}
\frac{1}{3\,!}\left[ \,x\frac{\partial \nu }{\partial x}(0,0,0)+y\frac{%
\partial \nu }{\partial y}(0,0,0)+z\frac{\partial \nu }{\partial z}(0,0,0)\,%
\right] ^{\,(3)}
\end{equation*}%
is not taken into account, and in  the  expansion
\begin{equation*}
\frac{1}{4\,!}\left[ \,x\frac{\partial \nu }{\partial x}(0,0,0)+y\frac{%
\partial \nu }{\partial y}(0,0,0)+z\frac{\partial \nu }{\partial z}(0,0,0)\,%
\right] ^{\,(4)},
\end{equation*}%
only the two terms
\begin{eqnarray*}
&&\frac{1}{4\,!}\left[ \,x^{4}\frac{\partial ^{4}\nu }{\partial x^{4}}%
(0,0,0)+y^{4}\frac{\partial ^{4}\nu }{\partial y^{4}}(0,0,0)+z^{4}\frac{%
\partial ^{4}\nu }{\partial z^{4}}(0,0,0)\right] ,  \\
&&\frac{6}{4\,!}\left[ \,x^{2}y^{2}\frac{\partial ^{4}\nu }{\partial
x^{2}\partial y^{2}}(0,0,0)
 +y^{2}z^{2}\frac{\partial ^{4}\nu }{\partial
y^{2}\partial z^{2}}(0,0,0)+z^{2}x^{2}\frac{\partial ^{4}\nu }{\partial
z^{2}\partial x^{2}}(0,0,0)\right]
\end{eqnarray*}
must be considered.
Taking   into account
\begin{eqnarray*}
&&\iiint_{\mathcal{D}}x^{2}y^{2}\varphi (r)dv=\iiint_{\mathcal{D}}y^{2}z^{2}\varphi
(r)\,dv\\
&&=\iiint_{\mathcal{\mathcal{D}}}z^{2}x^{2}\varphi (r)\,dv=\frac{4\,\pi }{15}\int_{\sigma
_{l}}^{\infty }\varphi (r)\,r^{6}\,dr,
\end{eqnarray*}%
and
\begin{eqnarray*}
&&\iiint_{\mathcal{D}}x^{4}\varphi (r)\,dv=\iiint_{\mathcal{D}}y^{4}\varphi
(r)\,dv\\
&&=\iiint_{\mathcal{D}}z^{4}\varphi (r)\,dv=\frac{4\,\pi }{5}\int_{\sigma
_{l}}^{\infty }\varphi (r)\,r^{6}\,dr,
\end{eqnarray*}%
we obtain
\begin{eqnarray*}
&&\iiint_{\mathcal{D}} \frac{1}{4\,!}\left[ \,x\frac{\partial \nu }{\partial x}(0,0,0)+y%
\frac{\partial \nu }{\partial y}(0,0,0)+z\frac{\partial \nu }{\partial z}%
(0,0,0)\,\right] ^{\,(4)}\varphi (r)dv \\
&&=\frac{\pi }{30}\int_{\sigma _{l}}^{\infty }\Delta
(\Delta\rho)\,\varphi (r)\,r^{6}\,dr.
\end{eqnarray*}
This means that we have to add to the second member of Eq. (\ref{2grad}) the term $c^4 \rho \,\Delta^2 \rho$, and
\begin{equation*}
W=\iiint_{\mathcal{D}}k\left[ \rho ^{2}+b^{2}\rho \,\Delta\rho + c^4 \rho \,\Delta^2 \rho\right] dv , \label{4grad}
\end{equation*}
where $\Delta^2 \rho \equiv \Delta (\Delta\rho)$  is the biharmonic operator and
\begin{equation}
2\,k\,c^4 = \frac{\pi }{30}\int_{\sigma _{l}}^{\infty}\varphi (r)\,r^{6}\,dr.  \label{tau}
\end{equation}
Consequently, for the specific internal energy and the volume free energy of the fluid in the fourth-gradient order case,
\begin{equation*}
 \varepsilon = \alpha(\rho,\eta)- \frac{\lambda}{2}\, \Delta \rho - \frac{\gamma%
}{2}\, \Delta^2 \rho,
\end{equation*}
and
\begin{equation*}
 F= f  (\rho,T) - \frac{\lambda}{2}\,\rho\, \Delta\rho - \frac{\gamma%
}{2}\, \rho\,\Delta^2 \rho,\label{gradquatre}
\end{equation*}
 respectively, with $\gamma= - 2\,k\,c^4$. Term $f (\rho,T)$ is the volume free energy of the homogeneous fluid of density $\rho$ and temperature $T$.

\subsection{\label{sec2.2} Numerical application to London's forces}
The London potential of fluid/fluid interaction usually writes
$\varphi _{ll}=-{c_{ll}}/{r^{6}}
$, where $c_{ll}$ is the intermolecular  coefficient (\cite{Israel,Hamak}). In the simple form $1/r^7$, the force range is infinite; to obtain a convergent expansion of the density in the volume integrals at   boundary
$\infty$, it is necessary to give   range $L$ of London's forces.
 London's forces being infinite for $r < \sigma_l$, we additively assume that they are null for $r >L$. Then
$ \varphi _{ll}=-c_{ll}/r^{6}$  when $\sigma _{l}< r < L$ and $ \varphi _{ll}=\infty$ when $r\leq \sigma _{l}$.
From Eqs.  (\ref{lambda}) and (\ref{tau}), we deduce
\begin{equation*}
\lambda =\frac{2\pi c_{ll}}{3\, m_l^2 \sigma_l},\quad  \gamma= \frac{\pi c_{ll}L}{30\, m_l^2},\end{equation*}
(when $\sigma_l/L\ll 1$, $L$ can be considered as infinite for the calculation of $\lambda$).

In the case of water, the physical measurements are  indicated in   $\textbf{c.g.s.}$ units \cite{Israel,Handbook,Gouin2}:
$ c_{ll} = 1.4 \times 10^{-58},\ \sigma_l = 2.8 \times 10^{-8},\ m_l = 2.99
\times 10^{-23},\ \lambda= 1.17 \times 10^{-5}$. The development of energy at the fourth order needs to take the force range into account; when $L = 2 \times 10^{-6}$ cm, which is the average range of van der Waals
forces, we get $\gamma = 3.28 \times 10^{-20}$. Let us note that if we slightly change  the $L$-value, the $\gamma$-value   changes only linearly.

The ratio $\gamma/\lambda$ has the dimension of a square-length  and $\sqrt{%
\gamma/\lambda}= 7.5 \times 10^{-8} $ cm.
Consequently, an appropriate unit of length at the molecular scale is $\ell =  \sqrt{%
\gamma/\lambda}$, which is of the same order than the molecular diameter.

\section{\label{sec3} Equation of isothermal conservative motions}

The principle of virtual works (or virtual powers) is always a convenient way to obtain the equation of
motions \cite{Lin,Seliger,Serrin,Gavrilyuk}. A particle is identified in a Lagrange
representation by reference position $\boldsymbol{X}$ of coordinates $X,Y,Z$
in   reference configuration $\mathcal{D}_{0}$; its position is given in $\mathcal{%
D}$ by the Euler representation $\boldsymbol{x}$ of coordinates $x,y,z$. The
variations of particle motions can be deduced from families of virtual
motions of the fluid written as
\begin{equation*}
\boldsymbol{X}=\boldsymbol{\Psi }(\boldsymbol{x},t;\beta ),
\end{equation*}%
where $\beta $ denotes a real parameter defined in the vicinity of $0$ and   the real motion corresponds to $\beta =0$. Virtual displacements in the reference configuration are
associated with any variation of the real motion and can be written as \cite{Gouin3},
\begin{equation*}
\delta  \boldsymbol{X}=\left. \frac{\partial \boldsymbol{\Psi }}{\partial
\beta } (\boldsymbol{x},t;\beta )\right\vert _{\beta =0}.
\label{displacement}
\end{equation*}
Variation $\delta  \boldsymbol{X}$ is \textit{dual} and
mathematically equivalent to Serrin's variation
 (\cite{Serrin}, p. 145). Neglecting the body forces, the Lagrangian of
the fluid writes,
\begin{equation*}
\Lambda= \frac{1}{2}\, \rho\,{\boldsymbol{u}}^{\star}\boldsymbol{u} - F ,
\end{equation*}%
where $\boldsymbol{u}$ denotes the   particle velocity and $^\star$ the transposition. The equation of isothermal motions
stationarises
\begin{equation*}
\mathcal{G}=\iiint_{\mathcal{D}}\Lambda~dv .
\end{equation*}
 The density
satisfies the   mass conservation
\begin{equation}
\frac{\partial\rho}{\partial t}+  \rm{div} (\rho \boldsymbol{u})   = 0\quad
\Longleftrightarrow\quad\rho \ \text{det}\, \boldsymbol{F} =\rho_{_0}(\boldsymbol{X}),
\label{density}
\end{equation}%
where $\boldsymbol{F}\equiv\partial \boldsymbol{x}/\partial \boldsymbol{X}$ and $\rho_{_0}$ is defined on $\mathcal{D}_{0}$. ù
Classical methods   yield  the variation
of $\mathcal{G}$.
The variation in $\mathcal{D}_{0}$ commutes with the derivatives with respect to $\boldsymbol{x}$
\ ($\delta {\rm{grad}}^{p}\rho = {\rm{grad}}^{p}\delta \rho,\ p\in N $). As usual, we assume that  virtual displacements are null
on   boundary  $\partial \mathcal{D}$ and consequently  variations of integrated terms
are null on this boundary.  By using Stokes'
formula, we integrate by parts; from $\delta \mathcal{G}=\mathcal{G}^{\prime }(\beta )|_{{%
\beta =0}}, $ we get  (see Appendix for details)
\begin{equation*}
 \delta \mathcal{G} = \iiint_{\mathcal{D}}\left\{ \left[\; \frac{1}{2}\,  %
\boldsymbol{u}^{\star}\boldsymbol{u}- \frac{\partial f(\rho,T)}{\partial \rho} + \lambda \, \Delta\rho+  \gamma  \, \Delta^2\rho \right] \delta \rho
   + \rho\, \boldsymbol{u}^{\star}\delta \boldsymbol{u}\right\} ~dv .
\end{equation*}
By taking Eq. (\ref{density}) into account,
\begin{equation*}
\delta \rho =\rho\  {\rm{div}}_{0} \,\delta \boldsymbol{X}+\frac{1}{\text{det}
\boldsymbol{F}}\frac{\partial \rho _{_0}}{\partial \boldsymbol{X}}\,\delta
\boldsymbol{X},
\end{equation*}%
where $\rm{div}_{0}$ is the divergence operator in $\mathcal{D}_{0}$.
The definition of the velocity implies
\begin{equation*}
\frac{\partial \boldsymbol{X}\,(\boldsymbol{x},t)}{\partial \boldsymbol{x}}%
\, \boldsymbol{u}+\frac{\partial \boldsymbol{X}\, (\boldsymbol{x},t)}{%
\partial t}=0,
\end{equation*}%
and therefore,
\begin{equation*}
\frac{\partial \delta \boldsymbol{X}}{\partial \boldsymbol{x}}\ \boldsymbol{u%
}+\frac{\partial \boldsymbol{X}}{\partial \boldsymbol{x}}\ \delta
\boldsymbol{u}+\frac{\partial \delta \boldsymbol{X}}{\partial t}=0\quad
\Longleftrightarrow\quad \delta \boldsymbol{u}=-\boldsymbol{F} \overset{ {\ \centerdot }}{%
 \widehat{\delta \boldsymbol{X}}},
\end{equation*}%
where the superposed dot is the material
derivative. By denoting
\begin{equation*}
 K=   \frac{\partial f(\rho,T)}{\partial \rho} -\lambda\, \Delta\rho - \gamma\, \Delta^2\rho  \quad {\rm and}\quad m=\frac{1}{2}\,{\boldsymbol{u}}^{\star}%
\boldsymbol{u}-K\,,%
\end{equation*}%
\begin{equation*}
\delta \mathcal{G} =\int \int \int_{\mathcal{D}}\left[ m~\delta \rho -\rho \, {( {%
\boldsymbol{u}^{\star}\boldsymbol{F}})}\,\overset{ {\ \centerdot }}{\widehat{%
\delta \boldsymbol{X}}}\right] ~dv
\end{equation*}
and by integration by part on $\mathcal{D}_0$,
\begin{equation*}
\delta \mathcal{G}= \int \int\int_{{\mathcal{D}}_{0}}\rho _{0}\left[ \overset{%
\centerdot }{(\widehat {\boldsymbol{u}^\star \boldsymbol{F}})}-{\rm{grad}}_{0}^{\star}\,m
\right] \delta \boldsymbol{X}~dv_{0},
\end{equation*}%
where ${\rm{grad}}_{0}$ is the gradient  and $%
dv_0$ the volume element,  in ${\mathcal{D}}_{0}$.

The principle of virtual work reads:
\begin{equation*}
For\ any\ displacement\  \delta
\boldsymbol{X} \ null\ on\ the\ edge\ of\  \mathcal{D}_{0},\
\delta \mathcal{G}=0.
\end{equation*}
We get $\  \overset{\centerdot }{(\widehat{%
\boldsymbol{u}^{\star}\boldsymbol{F}})}= {\rm{grad}}_{0}^{\star}\,m. $\ \
Noticing that
$$\left(\boldsymbol{a}^{\star}+\displaystyle\boldsymbol{u}^{\star}\frac{%
\partial \boldsymbol{u}}{\partial \boldsymbol{x}}\right)\boldsymbol{F}=\overset{%
\centerdot }{(\widehat{\boldsymbol{u}^{\star}\boldsymbol{F}})},$$
where $%
\boldsymbol{a}$
is the acceleration vector,
\begin{equation}
\boldsymbol{a}+{\rm{grad}}\, K = 0.  \label{motion}
\end{equation}%
Obviously,  $K\,$ has the
same physical  dimension as a chemical potential.
From $ \partial f(\rho,T) /{\partial \rho}\equiv \mu_{_{0}} (\rho,T)$, where $\mu_{_{0}}(\rho,T)$ is the chemical potential of the fluid  bulk, at equilibrium and temperature $T$  we get,
\begin{equation*}
K =\mu_{_{0}}   -\lambda\, \Delta\rho - \gamma\, \Delta^2\rho,  \label{CP}
\end{equation*}
and   equation of motion (\ref{motion}) yields the potential acceleration value
for the fourth-gradient fluid.
\section{\label{sec4} Case   of planar interfaces}
\subsection{Fluid at equilibrium}

The  one-dimensional equilibrium equation writes
\begin{equation}
K=  \mu_{_{01}},\quad{\rm with}\quad  K =
 \mu_{_{0}}  (\rho,T)  -\lambda \,\frac{d^2\rho}{dx^2}- \gamma \,\frac{d^4\rho}{dx^4} \label{densityequation}
\end{equation}
where  $d/dx$ is  the derivative with respect to  space variable
$x$ across the interface and $\mu_{_{01}}$ is a convenient
additive constant. If we consider the limit case when $\gamma =0$,
we are back to the Cahn and Hilliard  equation.
\\
Equation  (\ref{densityequation})  multiplied by $d\rho/dx$ yields,
\begin{equation*}
\frac{d}{dx}\left[\frac{1}{2}\,\lambda  \left(\frac{d\rho}{dx}\right)^2+
\int_{-\infty}^x  \gamma\,\frac{d^4\rho}{dx^4}\, \frac{d\rho}{dx}\, dx
\right]= \frac{d}{dx}\left[f  -\mu_{_{01}}\,  \rho\,
\right] .  \label{FirstIntegral}
\end{equation*}
Taking into account  the fact that
\begin{equation*}
\frac{d}{dx}\left[\int_{-\infty}^x \,\frac{d^4\rho}{dx^4}\, \frac{d\rho}{dx}\,
dx \right] = \frac{d}{dx}\left[\frac{d^3\rho}{dx^3}\, \frac{d\rho}{dx} - \frac{1}{2}%
\, \left(\frac{d^2\rho}{dx^2}\right)^2\right],
\end{equation*}
the equilibrium equation  has a first integral in the
form of an energy equation,
\begin{equation*}
\frac{1}{2}\,\lambda \left(\frac{d\rho}{dx}\right)^2+
  \gamma\left(\frac{d^3\rho}{dx^3}\, \frac{d\rho}{dx} - \frac{1}{2}%
\, \left(\frac{d^2\rho}{dx^2}\right)^2\right)= f   - \mu_{_{01}}\,\rho -f_{_{1}} ,
\label{FirstIntegral2}
\end{equation*}
where $f_{_{1}}$ is an additive constant.

\subsection{  Equation of one-dimensional travelling waves}
Let us study the  problem when   scalar velocity $u$ and
density $\rho$ are only functions of  variable $\zeta \equiv x - c\,t$, where   $t$ is the time and $c$ the wave celerity with
respect to a Galilean reference frame,
\begin{equation*}
u = u\,(x - c\,t), \ \ \rho =\rho\,(x - c\,t).
\end{equation*}
Mass balance equation \eqref{density} yields
\begin{equation*}
-c\, \frac{d\rho}{d\zeta} +\frac{d(\rho\, u)}{d\zeta} = 0
\end{equation*}
and by integrating, we obtain
\begin{equation*}
\rho\,(u-c) = q  \,,\label{massbalance}
\end{equation*}
where $q$ is constant along the interfacial motion. In the case  of waves, we obtain the
acceleration,
\begin{equation}
{a} = \frac{1}{2}\,\frac{d(u-c)^2}{d\zeta}= \frac{1}{2}\,\frac{d}{d\zeta}%
\left(\frac{q}{\rho}\right)^2. \label{acceleration}
\end{equation}
In  one-dimensional cases, Eqs.  (\ref{motion}) and  (\ref{acceleration}) yield
\begin{equation*}
\frac{d}{d\zeta}\left[\frac{1}{2} \left(\frac{q}{\rho}\right)^2+
\mu_{_{0}} -{\lambda} \,\frac{d^2\rho}{d\zeta^2}- \gamma \,\frac{d^4\rho}{d\zeta^4}\right]=0
\end{equation*}
and consequently,
\begin{equation}
 \lambda\,\frac{d^2\rho}{d\zeta^2}+ \gamma \,\frac{d^4\rho}{d\zeta^4} = \mu_{_{0}}  -\mu_{_{02}}+ \frac{1}{2} \left(\frac{%
q}{\rho}\right)^2,  \label{wavequ}
\end{equation}
where $\mu_{_{02}}$ is a convenient additive constant.

\subsection{Fluid at equilibrium near the critical point}

Near  the critical point, $\mu_{_{0}}  (\rho,T) -\mu_{_{01}}$ may be expanded in powers of $\varrho\equiv\rho-\rho_c$,  where $\varrho$ denotes the deviation of $\rho$   from its critical values $\rho_c$ :
\begin{equation}
\mu_{_{0}}(\rho,T) - \mu_{_{01}} = -A (T_c-T)\,\varrho +B\, \varrho^3, \label{expch}
\end{equation}
where A and B are two positive constants (\cite{rowlinson}, page 250) such that $A = \mu_{,_{11}}^c $ and $B =({1}/{6})\,\mu_{,_{30}}^c$ where
\begin{equation}
\mu_{,_{ij}}^c =   \frac{\partial^{i+j}\mu_{_{0}}} {\partial\rho^i\partial T^j} (\rho_c,T_c) \label{partialD}
\end{equation}
and $T_c$ is the critical temperature.\newline
When $T < T_c$\,, the densities $\rho_l$ and $\rho_g$ of the liquid and vapour  bulks  at phase equilibrium  satisfy the  Maxwell equal-area  rule, and  by symmetry  are the zeros  other  than $\rho=\rho_c$ in equation $\mu_{_{0}}  (\rho,T) -\mu_{_{01}} = 0$ :
\begin{equation*}
 \rho_{l} \sim \rho_c + \sqrt{ \frac{(T_c-T)}{B} }\ , \qquad \rho_{g} \sim \rho_c -\sqrt{ \frac{(T_c-T)}{B} }\ .
\end{equation*}
By taking    Eqs. (\ref{densityequation}-\ref{expch})   into account, we obtain at equilibrium
\begin{equation}
 \gamma \, \frac{d^4\varrho}{dx^4} + {\lambda} \,\frac{d^2\varrho}{dx^2}  =   -\,A \,(T_c-T)\,\varrho +B\, \varrho^3.\label{densityequationSH}
\end{equation}

\subsubsection{Rescaling of Eq. (\ref{densityequationSH})}
We  mainly consider the case when $T < T_c$ corresponding to a fluid   at a temperature lower than its critical value. One way to study Eq. (\ref{densityequationSH}) is as follows.
Two characteristic lengths are in competition :
\begin{equation}
\ell_1 = \sqrt{ \frac{\lambda}{ \, A\,(T_c-T)}}\ \quad  {\rm and} \quad \ell_2 = \sqrt[\textbf{4}]{ \frac{\gamma}{ \, A\,(T_c-T)}}\, .\label{lengths}
\end{equation}
 The two lengths are connected by the relation $d \times \ell_1 = \ell_2^2$\ \ with\ \ $d = \sqrt{\gamma/\lambda}$\,.  We define  scalar $\tau$   and mass $m$ as
 \begin{equation}
 \tau = \frac{\ell_2^2}{d^2} \equiv \frac{\lambda}{\sqrt{\gamma \,A\, (T_c-T)}}\,,\qquad m = \ell_2\,  \sqrt{\frac{\gamma}{B}}  \label{parameters}.
\end{equation}
In the  system of units associated with unit length  $\ell_2$  and unit mass  $m$, Eq. (\ref{densityequationSH}) writes
\begin{equation}
   \frac{d^4\varrho}{dx^4}+ \tau \, \frac{d^2\varrho}{dx^2} =    \varrho^3- \varrho,\label{densityequationSH1}
\end{equation}
which is an \emph{extended  form} of the \emph{Fisher-Kolmogorov equation}.
Let us consider the van der Waals equation of state,
\begin{equation*}
p = \frac{(R/M)\, T\,\rho}{1-b\,\rho}-a\,\rho^2 ,
\end{equation*}
where $p$ is the thermodynamical pressure of bulks, $R/M$ is the perfect gas constant  per unit mass, $a$ is the measure of   attraction between  particles and $b$ is the covolume previously defined in Section \ref{sec2.1}.
Classical   calculations using Eq. (\ref{partialD}) yield the values of $A$ and $B$:
\begin{equation}
A\equiv \mu_{,_{11}}^c = 6\,  \frac{p_c}{T_c\,\rho_c^2}  \,,\qquad B \equiv\frac{1}{6}\,\mu_{,_{30}}^c = \frac{3\,p_c}{2\,\rho_c^4}\, ,\label{AB}
\end{equation}
where $p_c$ is the fluid's critical pressure.
\newline
Let us note that when $T > T_c$, Eqs. (\ref{lengths},\ref{parameters}) are modified by replacing $(T_c- T)$ by $(T-T_c)$ and Eq. (\ref{densityequationSH1}) by
\begin{equation}
   \frac{d^4\varrho}{dx^4}+ \tau \, \frac{d^2\varrho}{dx^2} =    \varrho^3 + \varrho . \label{densityequationSH1'}
\end{equation}
\subsubsection{The case of water}
\begin{figure}[h]
\begin{center}
\includegraphics[width=7cm]{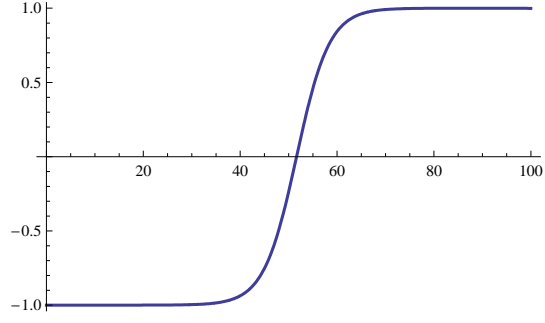}
\end{center}
\caption{Phase-transition kink $\left(\,{\rm in\ form}\ \varrho =  \mathcal F_{_0}(x)\,\right)$  near the water critical point   at $T = 646 {^{\,\circ}}{\rm K}$.   The units    associated  with length and mass in Eq. \eqref{densityequationSH1} are  such that     $x$-axis unit  is $2.7 \times 10^{-7}$ cm $\equiv 2.7$ nm  and $y$-axis unit  is $3.2 \times 10^{-6}$  g/cm$^{3}$.
}
\label{fig1}
\end{figure}

In the case of water, the physical measurements are  indicated in   $\textbf{c.g.s.}$ units \cite{Israel,Handbook,Gouin2} :
\begin{equation*}
 \rho_c   =   0.325, \quad p_c= 2,185 \times 10^{8},\quad T_c =   647.3 \ {^{\,\circ}}{\rm K}.
\end{equation*}
  From water values already proposed in Section \ref{sec2.2}, we get   $d = 5.3 \times 10^{-8} $ cm.  From Eq. (\ref{AB}), we obtain
 \begin{equation*}
\ell_1 \simeq   \frac{3 \times 10^{-8}}{\sqrt{1-(T/T_c)}}\  {\rm cm} \quad   {\rm and} \quad \ell_2 \simeq   \frac{4 \times 10^{-8}}{\sqrt[\textbf{4}]{1-(T/T_c)}}\ {\rm cm}\, .
\end{equation*}
Two lengths $\ell_1$ and $\ell_2$ are equal when $\tau=1$ corresponding to  temperature $T \simeq 430 {^{\,\circ}}\textrm{K}$, but for $T \simeq 646 {^{\,\circ}}{\rm K}$, $\tau\simeq 23$ and for $T \simeq 647.2 {^{\,\circ}}{\rm K}$, $\tau\simeq 46$ ;  only very close to the critical point,   $ {d^4\varrho}/{dx^4}$  can be neglected with respect to $\tau\,  {d^2\varrho}/{dx^2}$. Following the values of $\tau$, we can refer to  the monograph by  Peletier and Troy \cite{Peletier} to study the solutions of  Eqs. (\ref{densityequationSH1}) and  (\ref{densityequationSH1'}).
\newline
In Fig. \ref{fig1}  we draw the phase-transition in mass density near the critical point at $T \simeq 646 {^{\,\circ}}{\rm K}$. The form  of this phase-transition  pulse is not essentially different from the case when $\gamma =0$.

\subsection{Travelling waves near the critical point}

When body forces are neglected, Eq. (\ref{wavequ}) writes at t=0
(we replace $x$ by $\zeta $ to obtain the wave's profile at any time $t$):
\begin{equation}
\lambda \frac{d^{2}\varrho}{dx^{2}}+\gamma \frac{d^{4}\varrho }{dx^{4}}=-A\,(T_c-T) \,\varrho  + B\,\varrho^{3}+\frac{1}{2}\left(\frac{q}{\rho}\right) ^{2} +\mu_{_{02}}.
\label{waves2}
\end{equation}
Another way to study Eq. (\ref{waves2}) in place of method in Section \ref{sec2.1} is as follows.\newline We consider  the case when $T < T_c$.\newline
In place of $\rho $ (or $\varrho$), $x$ and $q$, we use the non-dimensional variables $Y$, $%
z$ and $Q$ such that :
$$
\rho =\rho _{c}(1+\varepsilon Y),\quad x={\mathcal L}\,z,\quad q=\chi\,Q,\
$$
with
\begin{equation*}
\varepsilon ^{2}=\frac{A(T_{c}-T)}{B\rho _{c}^{2}},\  {\mathcal L}^{2}=\frac{%
 \lambda }{A(T_{c}-T)},\ \chi^{2}=2\frac{A^{\frac{3}{2}}}{{B}^{\frac{1}{2}}}\,
(T_{c}-T)^{\frac{3}{2}}.
\end{equation*}%
Then, Eq. (\ref{waves2}) yields
\begin{equation}
\frac{d^{2}Y}{dz^{2}}+\left( \frac{\gamma\, A\,(T_c-T)}{\lambda^2 }%
\right) \frac{d^{4}Y}{dz^{4}}
= Y^{3}-Y+\frac{Q^{2}}{(1+\varepsilon Y)^{2}}+\mu_{1}. \label{waves3}
\end{equation}%
But
\begin{equation*}
\frac{Q^{2}}{(1+\varepsilon Y)^{2}}=Q^{2}-2\,\varepsilon\, Q^{2}\,Y+3\,\varepsilon\,
^{2}\,Q^{2}\,Y^{2}+\emph{0}\,(\varepsilon ^{3})
\end{equation*}
Equation (\ref{waves3}) yields
\begin{equation}
 \frac{d^{2}Y}{dz^{2}}+\left( \frac{\gamma A(T_c-T)}{\lambda^2 }%
\right) \frac{d^{4}Y}{dz^{4}}  =
 (Y+\varepsilon ^{2}\,Q^{2})^{3}-Y(1+2\,\varepsilon\, Q^{2})+\mu_{2},  \label{onde}
\end{equation}
where $\mu_1$ and $\mu_2$ are two additive constants.
Let us consider the new change of variables:
\begin{equation*}
Y=-\varepsilon ^{2}\,Q^{2}+ \kappa\, R,\quad z=\frac{y}{\kappa},\ \text{%
with }\kappa ^{2}=1+2\,\varepsilon\, Q^{2},
\end{equation*}
Equation (\ref{onde}) yields
\begin{equation}
 \frac{d^{2}R}{dy^{2}}+\left(\kappa^2 \frac{\gamma\, A\,(T_c-T)}{ \lambda^2 }%
\right) \frac{d^{4}R}{dy^{4}}  =R^{3}-R\, +\, R_{_0}.\label{onde1}
\end{equation}
where $R_{_0}$ is constant.
Generally, when we are close to equilibrium,  we can consider that $\kappa\approx 1$ and  Eq. (\ref{onde1}) reduces to :
\begin{equation*} \label{waves6}
 \frac{d^{2}R}{dy^{2}}+\left(\frac{\gamma\, A\,(T_c-T)}{\lambda^2 }%
\right) \frac{d^{4}R}{dy^{4}} =R^{3}-R\, +  \, R_{_0}\,  .
\end{equation*}
Again, we obtain  an Extended Fisher-Kolmogorov equation and, when $T < T_c$\,, we have to estimate the value of positive term
${\gamma\, A\,(T_c-T)}/{\lambda^2}$ which tends to zero when $T$ tends to  $T_c$\,. (The case $T > T_c$ can be analogously deduced).

\begin{figure}[h]
\begin{center}
\includegraphics[width=8cm]{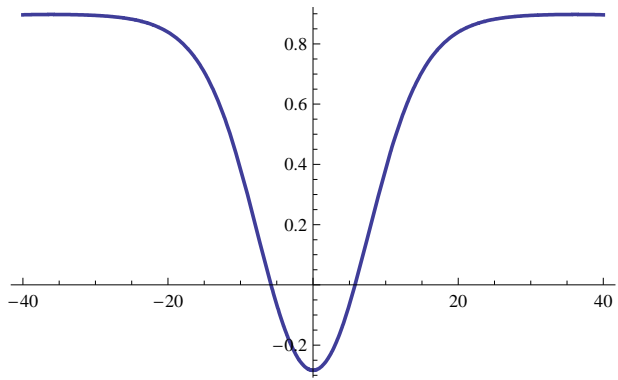}
\end{center}
\caption{Pulse $\left(\, {\rm in\ form}\ \varrho =  \mathcal F_{_1}(x)\,\right)$ near the water critical point at $T \simeq  646 {^{\,\circ}}{\rm K}$ when $\gamma = 0$ (second-gradient model);      $x$-axis unit  is $2.7 \times 10^{-7}$ cm $\equiv 2.7$ nm  and $y$-axis unit  is $3.2 \times 10^{-6}$  g/cm$^{3}$.
}
\label{fig2}
\end{figure}

\begin{figure}[h]
\begin{center}
\includegraphics[width=8cm]{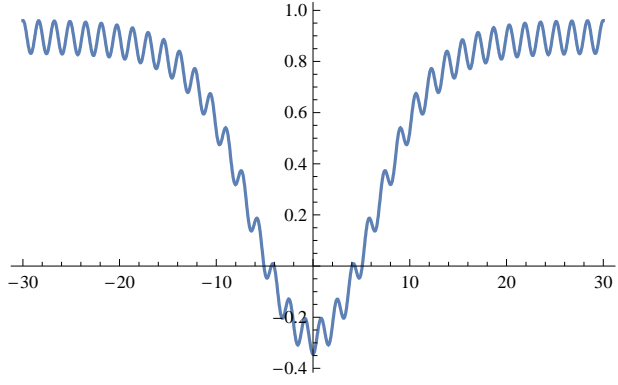}
\end{center}
\caption{Pulse $\left( \, {\rm in\ form}\ \varrho =  \mathcal F_{_2}(x)\,\right)$  near the water critical point at $T \simeq  646 {^{\,\circ}}{\rm K}$ when $\gamma \neq 0$ (fourth-gradient model); $x$-axis unit  is $2.7 \times 10^{-7}$ cm $\equiv 2.7$ nm  and $y$-axis unit  is $3.2 \times 10^{-6}$  g/cm$^{3}$.
}
\label{fig3}
\end{figure}

When the coefficient in front of the fourth-order derivative is positive,  solutions provided by the Extended Fisher-Kolmogorov equation are richer and more realistic than those provided by the Cahn-Hilliard equation in the vicinity of the critical point. Whereas the  Cahn-Hilliard theory predicts only the possibility of monotonic kinks, here different density profiles  are possible and especially oscillating pulse waves are admitted (cf. \cite{Peletier} - Chapter 9 and \cite{Chaparova}).  Figures \ref{fig2} and \ref{fig3} allow to compare - in   second and   fourth-gradient models - pulses near the critical point of water at $T \simeq 646 {^{\,\circ}}{\rm K}$.

\section{Concluding remarks}
Characteristic lengths $\ell_1$ and $\ell_2$, and   $\tau$ defined in Section \ref{sec3} grow  to infinity when $T$ tends to $T_c$.  Consequently, term $\tau\  d^2\varrho/dx^2$  is preponderant over term $d^4\varrho/dx^4$    and  Eq.  (\ref{densityequationSH1}) merges into
the corresponding Cahn-Hilliard equation. These results correspond  to  the point of view  in (\cite{Widom}, \cite{rowlinson} - Chapter 9) that near the critical point, the energy expansion of the fluid may be approximated by a gradient expansion typically truncated at the second order.
Nevertheless, the results we have found with an expansion truncated at the fourth order are more  in accordance with the  renormalisation group theory due to the possibility of oscillating pulse waves   generating  the density opalescence  observed as  the fluid  approaches its critical point  \cite{Widom2} and \cite{simulation}. The differences  in   pulse-wave oscillations between second and fourth-gradient models allow to revisit papers introducing kinks versus pulses as in \cite{Truskinovsky}. We believe that this result is remarkable and will hopefully stimulate further and deeper investigations on both  theoretical and phenomenological nature.
\newline
Finally, it is interesting to note - and it is not the case for the second gradient model - that the fourth-gradient model   is able   to take the range of London intermolecular forces into account.\newline\newline

\textbf{Appendix: }
{\bf{Some useful formulae}}\\

We take into account  the following results :
$$\rho\,  \rm{div}  {\rm{grad}} \,\rho=  \rm{div} (\rho\,{\rm{grad}}\,\rho) -({\rm{grad}}\, \rho)^2.$$
Term  $ \rm{div} (\rho\,{\rm{grad}}\,\rho)$ can be integrated on the boundary of $\mathcal{D}$, and consequently, $\displaystyle -\delta(\frac{\lambda}{2}  ({\rm{grad}}\,\rho)^2)$ corresponds in $\mathcal{D}$ to
$$
-\lambda\, {\rm{grad}}^{\star} \rho\  {\rm{grad}}\, \delta\rho \equiv -\lambda\, \rm{div} (\delta\rho \, {\rm{grad}}\, \rho) + \lambda\,  (\rm{div}{\rm{grad}} \rho)\,\delta\rho
$$
and the variation of $\ \displaystyle\frac{\lambda}{2}\,\rho\,\Delta\rho\ $ is  $\lambda\;\Delta\rho\,\delta\rho$.\newline

In a similar way,
$$\rho\,\rm{div} \big({\rm{grad}}(\rm{div} {\rm{grad}}\,\rho)\big)\equiv \rm{div} \big(\rho\,{\rm{grad}}(\rm{div} {\rm{grad}}\,\rho)\big)-{\rm{grad}}^{\star} \rho\ {\rm{grad}}(\rm{div} {\rm{grad}}\,\rho).
$$
Term $\rm{div} \big(\rho\ {\rm{grad}}(\rm{div} {\rm{grad}}\,\rho)\big)$ can be integrated on the boundary of $\mathcal{D}$ and
$$-\, {\rm{grad}}^{\star}\rho\ {\rm{grad}}(\rm{div} {\rm{grad}}\rho)\equiv-\rm{div} \big((\rm{div} {\rm{grad}} \,\rho) \  {\rm{grad}}\ \rho\big) + (\rm{div}  {\rm{grad}}\,\rho )^2. $$

Integrating on the boundary of $\mathcal{D}$ the term  $-\rm{div}  \big((\rm{div} {\rm{grad}}\,\rho)\, {\rm{grad}}\,\rho\big)$, and considering that the variation of $\rho\,\rm{div} \big({\rm{grad}}(\rm{div} {\rm{grad}}\,\rho)\big)$
is the same as the variation of $ (\rm{div} {\rm{grad}}\,\rho)^2$ we obtain
$$ 2\, (\rm{div}  {\rm{grad}}\,\rho)\,(\rm{div}  {\rm{grad}}\,\delta\rho)\equiv 2\; \rm{div} \big((\rm{div} {\rm{grad}}\,\rho)\,{\rm{grad}}\,\delta\rho\big)- 2\,{\rm{grad}}^{\star}(\rm{div} {\rm{grad}}\,\rho) \  {\rm{grad}}\,  \delta\rho.$$

Term $2\; \rm{div} \big((\rm{div}{\rm{grad}}\,\rho)\,{\rm{grad}}\,\delta\rho\big)$ can be integrated on the boundary of $\mathcal{D}$ and
$$- 2\,{\rm{grad}}^{\star}(\rm{div} {\rm{grad}}\,\rho)\, {\rm{grad}} \, \delta\rho \equiv - 2\,\rm{div} \big(\delta\rho\,{\rm{grad}}(\rm{div} {\rm{grad}}\rho)\big)+2\, \big(\rm{div} {\rm{grad}}(\rm{div}{\rm{grad}}\rho)\big) \delta\rho.$$

Term $- 2\,\rm{div} \big(\delta\rho\,{\rm{grad}}(\rm{div}{\rm{grad}}\rho)\big)$ can be integrated on the boundary of $\mathcal{D}$ and the variation of
$\displaystyle\frac{\gamma}{2}\,\rho\, \Delta^2\rho$ is    $$ \gamma\,\big(\rm{div} {\rm{grad}}(\rm{div} {\rm{grad}}\rho)\big)\,\delta\rho \equiv \gamma\, (\Delta^2\rho) \   \delta\rho.$$
 \\

\noindent \emph{\textbf{Acknowledgements:}}
\newline

\noindent\emph{H.G. and G.S. are grateful to the Carnot Star program at Aix-Marseille University and GNFM of Italian INDAM for partial supports. The authors thank Dr. M. G\u{a}r\u{a}jeu for his scrutiny in the numerical exactness of Fig. \ref{fig3}.}

\end{document}